\documentclass[oldversion]{aa}  

\usepackage{mathtools}
\usepackage{graphicx}
\usepackage{txfonts}
\usepackage{color}
\usepackage{pdftricks}
\begin{psinputs}
   \usepackage{pstricks}
   \usepackage{multido}
\end{psinputs}

\begin{document}

   \title{Multi-epoch monitoring of the AA~Tau like star V\,354~Mon}

   \subtitle{Indications for a low gas-to-dust ratio in the inner disk warp}

   \author{P. C. Schneider\inst{1,2}
          \and
          C. F Manara\inst{2,3}
          \and
          S. Facchini\inst{4}
          \and
          H. M. G\"unther\inst{5}
          \and
          G. J. Herczeg\inst{6}
          \and
          D. Fedele\inst{7}
          \and
          P. S. Teixeira\inst{8, 9,10}
          }

   \institute{
            Hamburger Sternwarte, Gojenbergsweg 112,
              21029 Hamburg, Germany,
                \email{astro@pcschneider.eu}
       \and       
       Scientific Support Office, Directorate of Science, European Space Research and Technology Center (ESA/ESTEC), Keplerlaan 1, 2201 AZ Noordwijk, The Netherlands
       \and 
       European Southern Observatory, Karl-Schwarzschild-Str. 2, D-85748 Garching bei M\"unchen, Germany
     \and  
     Max-Planck-Institut f\"ur extraterrestrische Physik, Gie\ss{}enbachstrasse 1, Garching bei M\"unchen
      \and
      Massachusetts Institute of Technology, Kavli Institute for Astrophysics and Space Research, 77 Massachusetts Avenue, Cambridge, MA 02139
      \and
      Kavli Institute for Astronomy and Astrophysics, Peking University, Yiheyuan 5, Haidian Qu, 100871 Beijing, People's Republic of China
      \and
      INAF - Osservatorio Astrofisico di Arcetri, Largo E. Fermi 5, 50125, Firenze - Italy
      \and
      Institut für Astrophysik, Universit\"at Wien, T\"urkenschanzstrasse 17,
1180 Vienna, Austria
\and
Institut de Ci\`encies del Cosmos (ICCUB), Universitat de Barcelona (IEEC-UB), Mart\'i Franqu\`es 1, E08028 Barcelona, Spain 
\and
Scottish Universities Physics Alliance (SUPA), School of Physics and Astronomy, University of St. Andrews, North Haugh, St. Andrews, Fife KY16 9SS, UK
}

   \date{received; accepted}

  \abstract
   {
       Disk warps around classical T~Tauri stars (CTTSs) can periodically obscure 
       the central star for some viewing geometries. 
       For these so-called AA~Tau-like variables, 
       the obscuring material is located in the inner disk and
       absorption 
       spectroscopy allows one to characterize its 
       dust and gas content. Since the observed emission from CTTSs consists of several components 
       (photospheric, accretion, jet, and disk emission), which can all vary with time, 
       it is generally challenging to disentangling
       disk features from emission variability.
       Multi-epoch, flux-calibrated, broadband spectra provide us with the
       necessary information to cleanly separate absorption from emission variability. We applied this method to
       three epochs of VLT/X-Shooter spectra of the CTTS V\,354~Mon (CSI Mon-660) located in NGC~2264  and 
       find that: (a) the accretion emission remains 
       virtually unchanged  between the three
       epochs; (b) the broadband flux evolution is best described by disk 
       material obscuring part of the 
       star, and (c) the Na and K gas absorption lines show only a minor increase 
       in equivalent width during phases of high dust extinction.
       The limits on the absorbing gas column densities indicate a low gas-to-dust ratio in the inner disk, less 
       than a tenth of the ISM value. We speculate  that the evolutionary state of 
       V\,354~Mon, rather old with a low accretion rate, is responsible 
       for the dust excess through an evolution toward a dust dominated disk or 
       through the fragmentation of larger bodies that drifted inward 
       from larger radii in a still gas dominated disk.
     }


   \maketitle
%

\section{Introduction}
Young stars are surrounded by copious amounts of circumstellar gas and dust. A 
fraction of this material survives the main accretion phase and can still surround
the young star after a few million years in the form of a circumstellar disk, which might
evolve into a debris disk dominated by reprocessed dust grains. The 
structure of protoplanetary disks is currently subject to intense observational and 
theoretical investigations, often focusing on intrinsic disk emission
\citep{Nomura_2016,Ansdell_2016,Cox_2017, Loomis_2017,Long_2017}.

\begin{figure}[t!]
\centering
\includegraphics[width=0.4\textwidth,trim=80.0 180.0mm 50mm 25mm, clip]{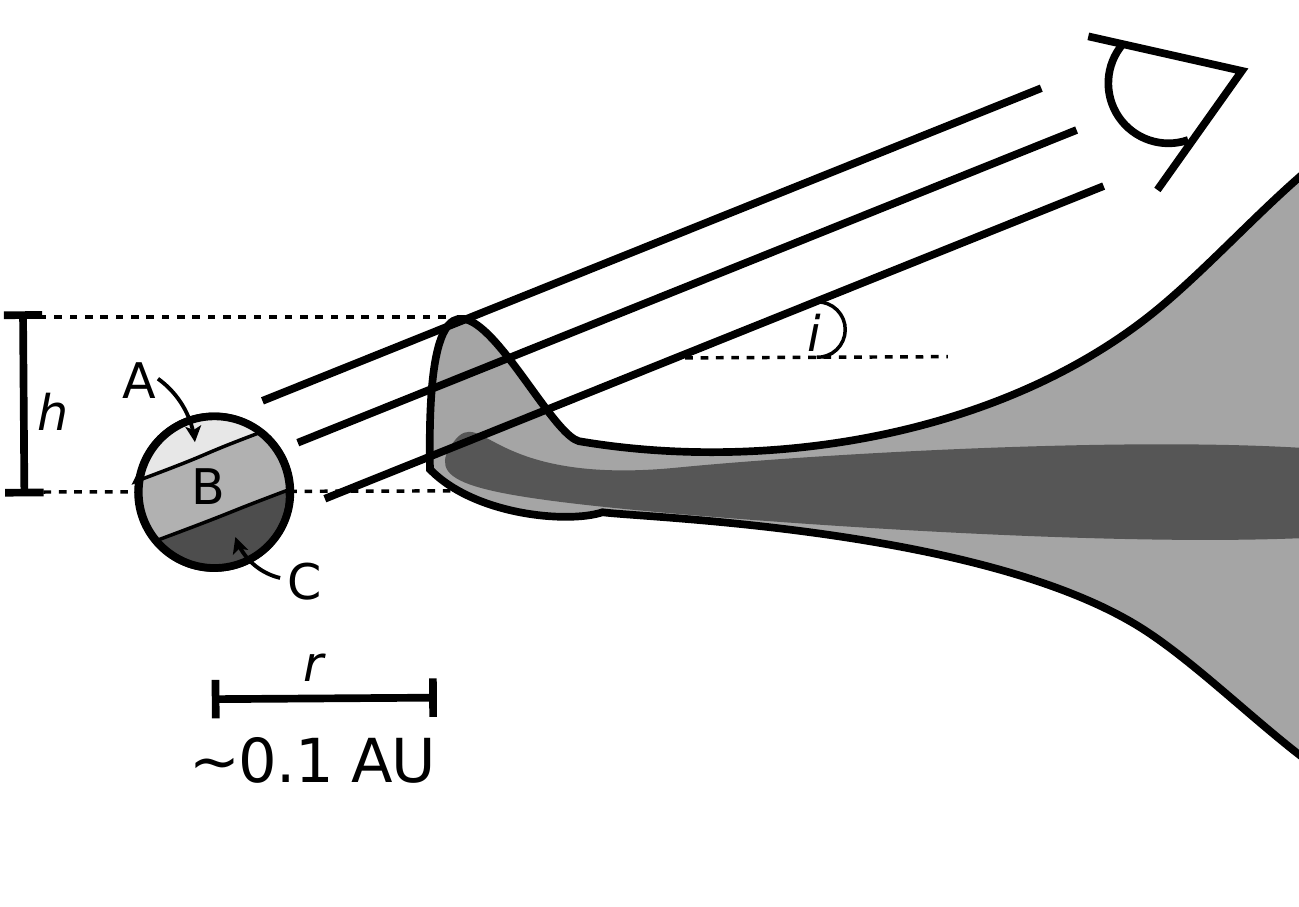}
\caption{Sketch of the inner disk structure around CTTSs indicating
our three component model. The unocculted part of the stellar disk is denoted with \emph{A}. The upper disk layer causes reddening of part the stellar disk (part \emph{B} of the stellar disk). Stellar emission from the region denoted with C is subject to opaque (gray) extinction by the lower disk layers. We also show the inner disk radius $r$, the viewing inclination $i$, and the height $h$ of the inner disk warp with respect to the disk midplane. \label{fig:sketch}}
\end{figure}

An alternative to these disk emission studies is absorption 
spectroscopy. So-called AA~Tau-like
variables show periodic dips in their light curves
 \citep[e.g.,][]{Bouvier_1999,Bouvier_2007, McGinnis_2015}.
These dips are thought to be caused by warps in the inner disk that rotate
between the star and the observer so that the line of sight periodically 
passes through the warp. Hence, differential absorption spectroscopy 
provides us with 
the properties of the obscuring material, i.e., the inner disk material
that is currently unaccessible by (interferometric) imaging techniques.
The periods of the dips are typically five to ten days, which translates to radii of about 0.1\,au
assuming Keplerian rotation. Hence, the warp material is close to or even 
cospatial with the inner disk edge.
Figure~\ref{fig:sketch} sketches the envisaged geometry, which
naturally results from the interaction between the stellar magnetic field 
and the inner disk edge \citep[][]{Foucart_2011}.

A related, perhaps overlapping, class of objects are the 
so-called dippers \citep{Stauffer_2015}. Their light curves bear many similarities to those of 
AA~Tau-like objects, but the dips are less pronounced and shallower than for those of ``classical'' 
AA~Tau-like objects. For both types of light curves, material at the inner disk edge is 
responsible for the dimmings. However, it is somewhat unclear if these two types of light curves
really describe two 
different phenomena (e.g., a persistent disk ``wall' vs. intermittent material in the disk atmosphere) or 
rather the same mechanism but of variable strength (e.g., variable filling of a warped inner disk atmosphere). Recent studies
show that the inner and outer disks are not necessarily aligned  
\citep{Ansdell_2016,Scaringi_2016} and absorption by inner disk material 
is also observed in systems with rather inclined outer disks, including the prototypical AA~Tau system, where ALMA observations of dust in the outer disk reveal a disk inclination of about $59^\circ$, meaning they are decidedly not close to edge-on \citep{Loomis_2017}.

Specifically for the \object{AA Tau} system, no strong color changes during the eclipses
have been found. In other words, the extinction is rather gray, which suggests that the extinction is caused by
large grains in the line of sight \citep{Bouvier_2003}.
However, accretion variability can also change the observed stellar 
spectrum. Due to disk-locking \citep[e.g.,][]{Matt_2010}, the stellar period and the 
rotation period of the inner disk warp are likely (almost) synchronized
so that it is challenging to disentangle accretion and extinction
properties using photometry only.

Further, the disk warps are not stable and changes in 
the inner disk geometry are the preferred explanation for major dimming 
events of CTTSs. Examples of these dimming events are AA~Tau \citep{Bouvier_2013, Schneider_2015},
\object{RW Aur} \citep{Petrov_2015, Schneider_2015_RW, Bozhinova_2016, Facchini_2016} as well a few other objects \citep{Hamilton_2001,Rodriguez_2015,Rodriguez_2017}. Combining the findings from these studies, which use information from X-ray to NIR wavelengths, shows that the dimming events are likely caused by perturbations in the inner disk with a scale height of about one stellar radius and that the material has a gas-to-dust ratio close to the ISM value.

Here, we present flux-calibrated 
X-Shooter spectra of the CTTS \object{V354 Mon} to measure simultaneously
(a) accretion signatures using line fluxes, (b) dust extinction 
through changes in the broadband spectral energy distribution 
(SED), and (c) changes in the absorbing gas column density from
inspection of strong absorption lines. 
Our paper is structured as follows. The target, V\,354~Mon, is described in Sect.~\ref{sect:target}. The X-Shooter observations are described in Sect.~\ref{sect:obs}
and we present our results in Sect.~\ref{sect:results}. In Sect.~\ref{sect:disc} we provide a
brief discussion of our results and we close with the conclusions in Sect.~\ref{sect:concl}.

\section{V\,354~Mon \label{sect:target}}
The CTTS V\,354~Mon (or CSI Mon-660) is
located in NGC~2264 \citep[d=760 pc,][]{Dahm_2008} and
has been extensively
studied in the framework of the CSI~NGC~2264 project \citep[e.g.,][]{Alencar_2010}.
Specifically, \citet{McGinnis_2015} classify V\,354~Mon as a stable AA~Tau-like 
variable with a period of 5.25 days where ``stable'' indicates that regular 
eclipses with a depth of about 1\,mag are present during both CoRoT observing epochs (2008 and 2011) of approximately one month length each.

\citet{Fonseca_2014} present a detailed study of V\,354~Mon 
using photometry and H$\alpha$ profiles covering about two rotation 
periods. They correlate the flux evolution with absorption features seen
against the H$\alpha$ line and find that redshifted absorption is most pronounced
during the epoch of the lowest flux. \citet{Fonseca_2014}
suggest that an accretion funnel causes the redshifted H$\alpha$ absorption
with the hotspot located under the funnel facing the observer during the 
phase of the flux minimum. Furthermore, they find that a hotspot 
cannot contribute significantly to the observed spectrum. Instead, they 
model the (approximately sinusoidal) evolution of the photometric magnitudes 
as variable circumstellar material, i.e. a disk warp,
which extends about $h/r=0.3$ above the disk midplane ($h$ and $r$ being 
the height above the disk midplane
and the inner disk radius, respectively). Here, we want 
to use exactly this property, the modulation of the observed 
stellar flux due to circumstellar material, to obtain further insight
into the composition of the disk material.

\section{Observations and data analysis \label{sect:obs}}
The star V\,354~Mon was selected based on the known variability in previous CoRoT observations.
Observations were performed over three consecutive nights in 2010 with X-Shooter
at airmasses between 1.21 and 1.27. 
Table~\ref{tab:data} provides an overview of the observations
and Fig.~\ref{fig:SEDs} shows the flux calibrated spectra. The spectral region covered 
by X-Shooter ($\lambda\lambda\sim$300-2500 nm) is observed simultaneously
and the spectra are divided into three arms, UVB ($\lambda\lambda\sim$ 300-550 nm), 
VIS ($\lambda\lambda\sim$ 550-1010 nm), and NIR ($\lambda\lambda\sim$ 1010-2500 nm).
Observations were carried out using the narrower slits in the VIS and NIR arms (0\farcs4~width) 
and the 1\arcsec~wide slit in the UVB arm. Before the exposures with the narrow slits, a 
short exposure with a slit width of 5\arcsec~width was performed in order 
to account for slit losses and to spectrophotometrically
calibrate the spectra. The spectra were acquired by nodding the telescope along a direction 
perpendicular to the slit and the nebular and sky emission was subtracted when combining the four nodding frames.
Data reduction follows the same procedure as in \citet{Manara_2016}. Briefly,
the spectra were reduced with the ESO X-Shooter pipeline version v.1.3.2  \citep{Modigliani_2010}. 
The 1D spectra were manually extracted from the rectified 2D spectra produced by the pipeline.
Flux calibration was performed using observations of the spectrophotometric standard stars
GD-71 (observed at the beginning of each night) and Hiltner-600 (observed every one to two hours each night). Finally, the narrow-slit spectra were rescaled
upward to match the large-slit spectra \citep[see][for a detailed discussion of the accuracy and stability of the flux calibration method]{Rugel_2017}. The VIS arm was corrected for telluric absorption lines 
using a telluric standard star observed close in time and airmass to the target.
We derived the stellar, heliocentric rest velocity as
$18.6 \pm 1.0$\,km\,s$^{-1}$ from fitting the Li$\,\lambda671$ line, which is compatible with 
the velocity of the NGC~2264 cloud \citep{Furesz_2006}. 

The X-Shooter spectra include atomic gas absorption lines, which we fit
with Voigt-profiles using PyAstronomy\footnote{\url{https://github.com/pcschneider/PyAstronomy}}. 
Specifically, we allowed the central wavelengths, line amplitudes and line widths 
to vary during the fit. We estimated the local error by the standard deviation in 
presumably line-free wavelength regions and provide 1~$\sigma$ errors. 
Further, we required the two members of the doublets under consideration 
to have the same width.
The 
equivalent width (EW) is derived from this fit to provide reasonable errors
(we used Markov chain Monte Carlo methods and $5\times10^4$ iterations).
These values agree well with the EW derived from direct integration.

Our analysis is differential and most values are measured with respect to 
the brightest spectrum (2010-01-20). To simplify the discussion, we have assumed in 
the following that the brightest spectrum corresponds to the uneclipsed star. 
If the brightest spectrum is already subject to partial circumstellar 
extinction, our conclusions change only in so far that the values should 
be corrected for the difference between the brightest spectrum and the 
uneclipsed star.

\begin{figure}[t]
\includegraphics[width=0.49\textwidth]{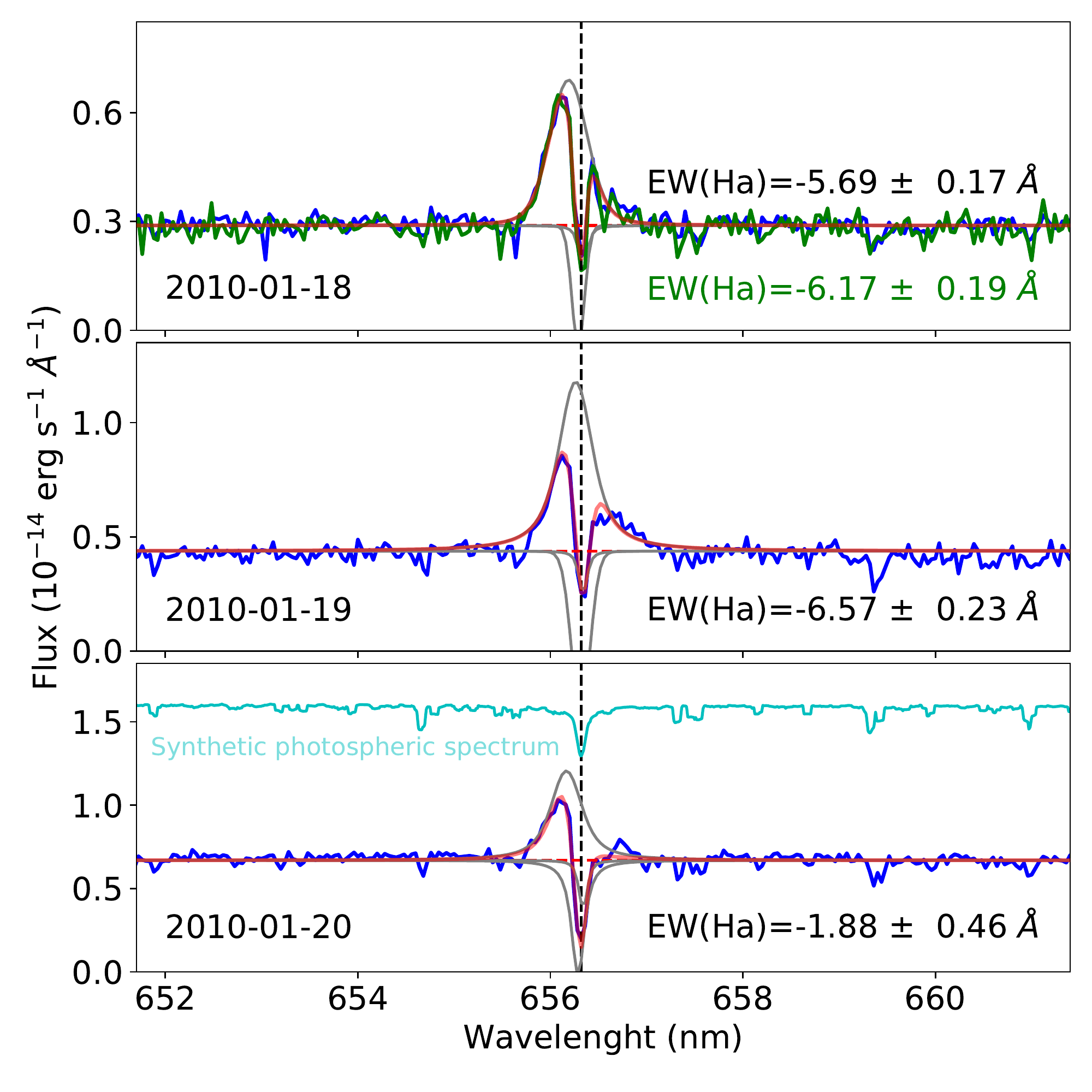}
\includegraphics[width=0.49\textwidth]{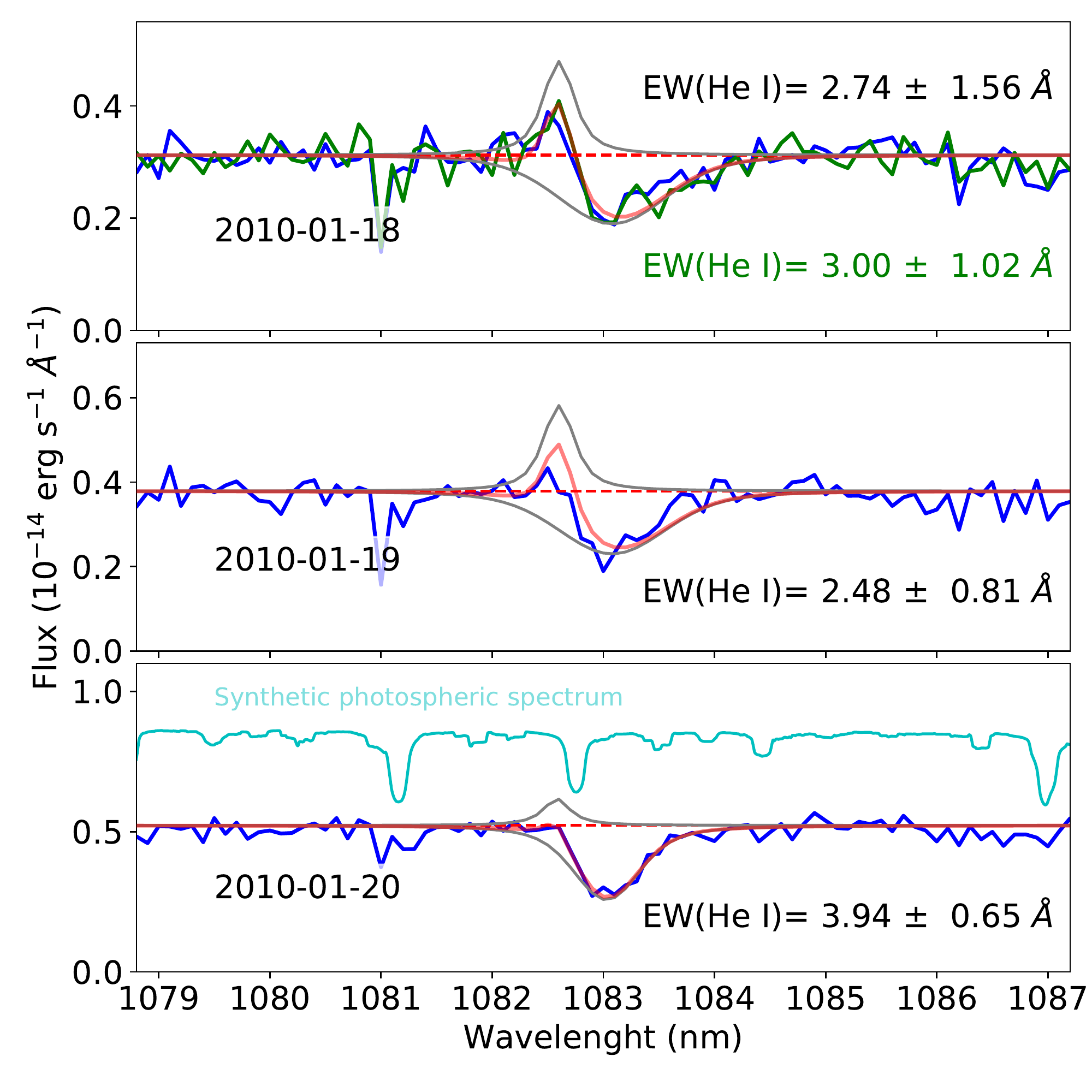}
\caption{{\bf Top}: H$\alpha$ spectra fitted with a two component
model (one emission, one absorption Voigt-profile) including a
stellar absorption component.
{\bf Bottom}: He~{\sc i} spectra fitted with a two  component
model (one emission, one absorption Voigt-profile).
Individual fit-components are shown in gray. The synthetic spectrum 
is offset for clarity.
          \label{fig:Ha}}
\end{figure}

\section{Results \label{sect:results}}

We first derived the stellar parameters and accretion properties
from the brightest spectrum (2010-01-20), which has a V~magnitude
that is within 0.1\,mag of the value
quoted in \citet{Fonseca_2014}, meaning that it is virtually equal to the uneclipsed spectrum.
The two other spectra are dimmer by 0.5~mag (2010-01-19) and 1.0~mag (2010-01-18) in the V band.
We note that extrapolating the CoRoT data\footnote{ ``Transit'' center: MJD = 54534.22 or JD=2454534.72, i.e., time of max. disk absorption; orbital period: 5.25\,days \citep[from][]{McGinnis_2015}.} places the 2010-01-18 spectrum close to flux maximum, which also explains the choice of integration times. This is  contrary to our data, but already a minor change in period of only 0.02\,days (fully compatible with the combined 2008+2011 CoRoT dataset) shifts the disk warp phases associated with the X-Shooter observing dates by half a period. Hence, the X-Shooter data are fully compatible with an unchanged disk warp period.

\subsection{Stellar parameters and accretion}
For reference, we started by fitting the brightest spectrum with a combination of synthetic spectra, accretion emission, and absorption using the methodology of \citet{Manara_2013, Alcala_2014}. We find 
a spectral type of K2 \citep[$T_{\rm eff}=4900\,$K, close to K4 derived by ][]{Fonseca_2014},  M$_\star\sim1.4-1.5\,M_\odot$ 
and a quite large age of $\sim7$\,Myr
for the \citet{Siess_2000} evolutionary tracks, which is somewhat older than the 
age of 3\,Myrs  derived by 
\citet{Lamm_2004} and \citet{Flaccomio_2006}. The corresponding stellar radius is about $1.8\,R_\odot$ while \citet{Fonseca_2014} estimate $R_\star = 2.4\,R_\odot$. The difference is mainly caused by a slightly lower $T_{eff}$ assumed by \citet{Fonseca_2014}. We used these values to provide phtospheric EWs for reference and to transfer orbital periods to radii, but our key diagnostics are
entirely differential between the three individual spectra so
that the exact values are of minor importance. 
Still, we note that the values for V\,354~Mon recently published by \citet{Venuti_2018} agree quite well with our values ($T_{eff} = 4836\,$K, $A_V$=0.76, $L_{bol} = 1.964\,L_\odot$, $M_\star = 1.59\,M_\odot$, EW(H$\alpha$) = 1.2\,\AA).
 
Assuming that the brightest spectrum is only subject to interstellar 
reddening we find $A_V=0.7$\,mag for $R_V=3.1$ and the \citet{Cardelli_1989}
reddening law.
The best fit requires some 
excess in the blue part of the spectrum,
which corresponds to $L_{excess}/L_\star\approx0.22\,L_\odot/1.67\,L_\odot\approx0.13$. Interpreting 
this excess as accretion would imply an accretion rate 
$\dot{M}_{acc}$ of about $2\times10^{-8}\,M_\odot$/yr, which is a typical value for 
CTTSs with the stellar mass of V\,354~Mon. However, the other epochs do not 
require excess blue emission, and emission line fluxes that usually 
go along with such accretion rates are not seen in any epoch. 
In fact, most of the accretion tracers do not have significant 
negative equivalent widths, let alone fluxes expected for an accretion rate of $\dot{M}_{acc} \sim 10^{-8}\,M_\odot$\,yr$^{-1}$. 
Only H$\alpha$ is significantly in emission during all epochs (Fig.~\ref{fig:Ha}, 
$EW=-7\dots-2$\,\AA), but its EW falls well short of the expection based on
an accretion rate of $10^{-8}\,M_\odot$\,yr$^{-1}$ \citep[H$\alpha$~EW expected $>100$\,\AA{}, see][]{Alcala_2014}. Furthermore,  
the H$\alpha$ EW is lowest during the brightest epoch 
\citep[consistent with the findings of ][]{Fonseca_2014}.

\begin{table}[t!]
\centering
  \caption{Observing details \label{tab:data} }
  \begin{tabular}{l l l r l l}
    \hline
    \hline
    Date & JD  & ObsID & \multicolumn{1}{c}{$t_{exp}$} & Airmass\\
    Jan. 2010     & (2455000+)& &  (s) &\\
    \hline
    18th 02:21:51 & 214.59 &  200198250 & 80   & 1.27 \\
    18th 02:32:23 & 214.60 & 200198252  & 50   & 1.26 \\
    19th 03:12:39 & 215.63 & 200198321  & 50   & 1.21 \\
    20th 02:40:46 & 216.62 & 200198391  & 120  & 1.23 \\
    \hline
  \end{tabular}
\end{table}

The H$\alpha$ profiles show a strong absorption feature close to the 
stellar rest velocity (Fig.~\ref{fig:Ha}).  Modeling the H$\alpha$ profile with two Voigt
profiles, this absorption component is redshifted with respect to the emission
component by about 24\,km\,s$^{-1}$ although the exact velocity difference
is somewhat uncertain due to the cross-talk between absorption and emission. 
There are no significant differences in the ratios between the amplitudes of the 
absorption and emission component. The flux in the emission component of 
the H$\alpha$ line ($\approx10^{30}\,$erg\,s$^{-1}$) points to accretion rates $\lesssim10^{-10}\,M_\odot$\,yr$^{-1}$ \citep[e.g.,][]{Dahm_2008, Alcala_2014}, 
compatible with the other emission lines. 
Also, the depth of photospheric lines
does not change between the epochs (see Fig.~\ref{fig:veiling}), 
which would be expected if strong accretion variability were present.
Therefore, we do not consider the blue 
excess in just one epoch as indicative of strong accretion, but rather as an artifact
of the fitting procedure given the potentially complex circumstellar structure.

We also inspected the He~{\sc i} lines (at 588\,nm and 1\,$\mu$m) in an attempt to 
correlate redshifted absorption and disk warp phase similar to the investigation by \citet{Fischer_2008} for AA~Tau. For V\,354~Mon, the redshifted absorption component of the He~{\sc i}~1\,$\mu$m line shows an approximately 10\,\% increase in the EW  between the dimmest and the brightest spectrum (see Table~\ref{tab:gas}) while the data for He~{\sc i}\,$\lambda588$\,nm is insufficient to draw solid conclusions on its EW evolution.  In any case, the observed changes are considerably smaller than those observed for AA~Tau by \citet[][]{Fischer_2008}, where the EW increased from 0.4 to 4.4\,\AA{} within a similar range of disk warp phases. Taken at face value, the V\,354~Mon trend also goes into the opposite direction, i.e., the largest EW pertains to the brightest spectrum contrary to the findings by \citet{Fischer_2008} for AA~Tau. We caution, however, that the minor changes observed for the He~{\sc i} EW in V\,354~Mon could be additionally affected by an interplay with the weak emission component that is 
spectrally not well resolved by X-Shooter. Therefore, we refrain from a detailed investigation noting only that the lack of a strong change in the He~{\sc i} absorption and the difference to AA~Tau is entirely compatible with the derived low accretion rate and the associated lack of strong absorption by the accretion funnel in He~{\sc i}.

In summary, signatures of strong accretion are not significantly 
detected; emission from the accretion 
shock is negligible in shaping the observed spectra compared 
to absorption effects, which agrees with the interpretation by 
\citet{Fonseca_2014}. Thus, we interpret the flux evolution observed
in the X-Shooter spectra
as an effect of variable circumstellar absorption.

\subsection{Dust extinction \label{sect:dust}}
\begin{figure*}[t!]
\centering
\includegraphics[width=0.9\textwidth]{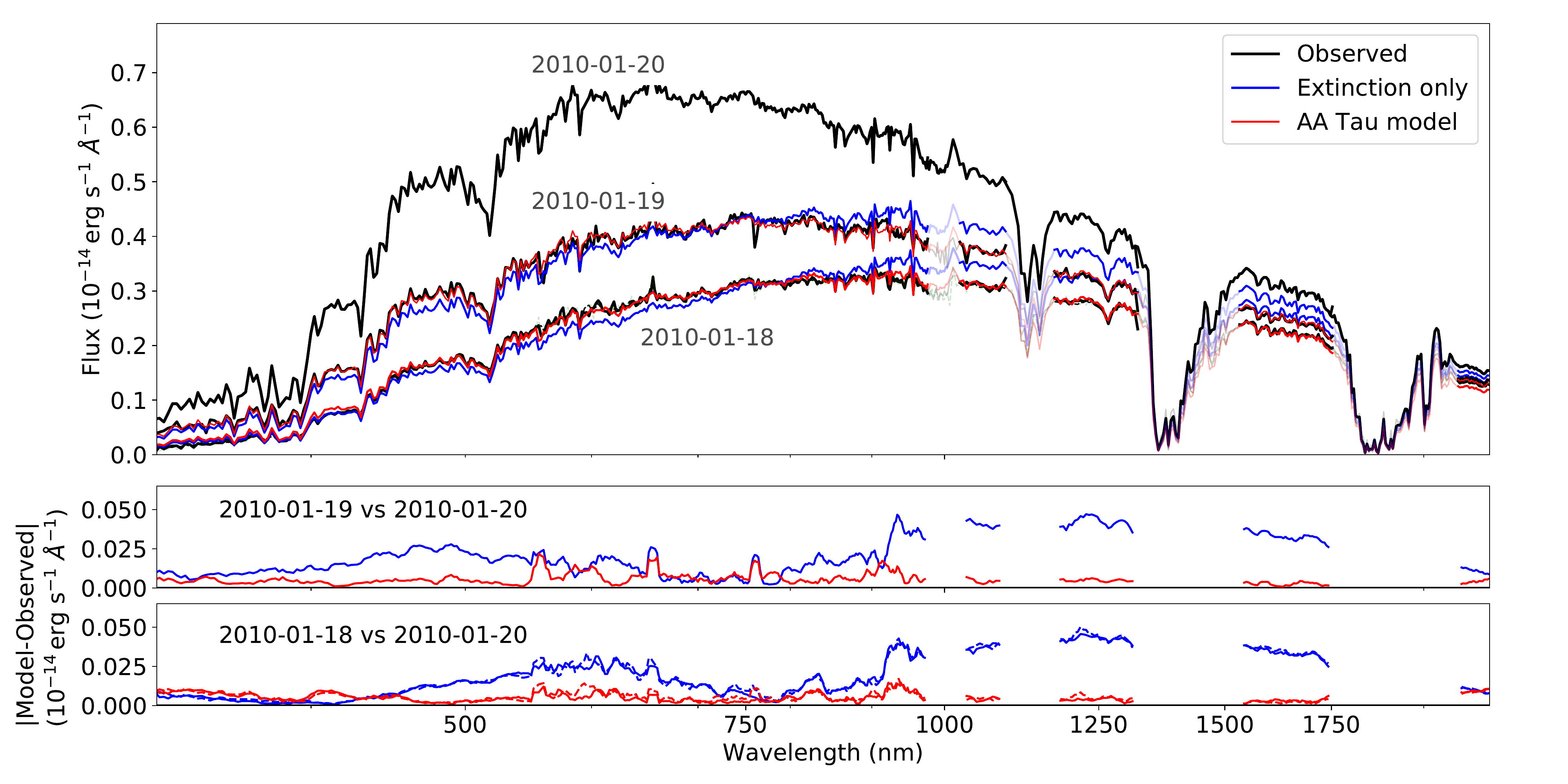}
\caption{{\bf Top}: Observed spectral energy distributions. 
         {\bf Bottom} two panels: Absolute difference between model and observed spectra.
         Models for the 
dimmer spectra (18th and 19th Jan.) assume that the difference
between the brightest spectrum (20th Jan.) and the dimmer
spectra is caused (1) only by additional extinction allowing $E(B-V)$
and $R_V$ to vary (blue), and (2) our AA~Tau model in red (see text for a detailed description).
To differentiate between the two 18th Jan. spectra, the shorter one
is displayed as a dashed line. Spectral ranges strongly affected by telluric absorption are 
removed from the fit and shown in lighter colors. The parameters corresponding to the 
blue and red curves in the bottom two panels are given in Table~\ref{tab:fit}. \label{fig:SEDs}}
\end{figure*}

Here we investigate the flux evolution 
and model it as
an extinction effect since there are no indications 
for veiling changes due to accretion variability. We find that changing only the extinction between the epochs cannot explain 
the differences between the spectra 
even when allowing $E(B-V)$ to vary freely and $R_V$ to 
vary between two and eight. The blue part of the spectrum would be
underpredicted while the flux at longer wavelengths would be overpredicted, meaning that the predicted SED for the dimmer spectra is too red compared 
the observed flux although we 
allow for exceptionally large values of $R_V$ (see Fig.~\ref{fig:SEDs}).

Extending the extinction-only model with a gray absorption component 
does not resolve the discrepancies. However, this model allows us 
to check the possibility of a significant scattering contribution as follows.
First, we restricted the fit to wavelengths longward of 8000\,\AA{} where 
scattering should be less important than at shorter wavelengths. Second, 
we performed a fit allowing $E(B-V)$, $R_V$ and the gray extinction part to
vary. While this model provides a reasonable fit to the red spectral 
evolution, it underpredicts the flux at blue wavelengths. This is exactly the 
part of the spectrum in which we would expect scattering to contribute.
However, about 15\,\% of the flux in the bright state would be required
to provide the missing flux. Such a high scattering efficiency is 
very unlikely given that a ``standard'' protoplanetary  disk intercepts only 
about 10\,\% of the stellar light, so we ignore scattering in the 
following. We note, however, that scattering can modify the derived values (slightly).
Most of the missing blue flux 
must come from a different source: the fraction of the 
star that is not eclipsed by the disk warp as detailed in the following.

\begin{table}[t!]
\centering
\setlength{\tabcolsep}{0.1cm}
\caption{Results of the spectral fitting. $A_V$ is derived as the change in V-band magnitude caused by the reddening and not part of the fitting procedure. \label{tab:fit}}
\vspace*{-0.2cm}
\begin{tabular}{ l c c}
\hline\hline
Parameter & \multicolumn{2}{c}{Value}\\
          & 2010-01-19 & 2010-01-18 \\[0.2cm]
      \hline
\multicolumn{3}{c}{Extinction only}\\
\multicolumn{3}{c}{(blue lines in the bottom two panels of Fig.~\ref{fig:SEDs})}\\[0.2cm]
$E(B-V)$ & 0.08 & 0.19 \\

$R_V$ & 8.0 & 6.0 \\
$A_V$ (derived) & 0.62 & 1.16\\[0.2cm]
\hline\\[-0.1cm]
\multicolumn{3}{c}{AA Tau model w/ $R_V=3.1$}\\
\multicolumn{3}{c}{(red lines in the bottom two panels of Fig.~\ref{fig:SEDs})}\\[0.2cm]

$E(B-V)$ & $0.69\pm0.11$ & $0.53\pm0.07$\\
$A_V$ &  0.45 & 0.96\\[0.2cm]

reddened ($S_B$) & $0.38\pm0.02$ & $0.64\pm0.07$\\
unobscured ($S_A$) & $0.54\pm0.02$ & $0.23\pm0.04$\\
\hline
\end{tabular}
\vspace*{-0.4cm}
\end{table}

Next, we considered a mathematical description for the flux evolution that represents
the physical model sketched in Fig.~\ref{fig:sketch},
which is based on the AA~Tau scenario \citep[e.g.,][]{Bouvier_2003, Schneider_2015}. 
We assumed that the absorption is caused by
a warp in the inner disk that grazes the stellar disk. The warp's linear dimensions 
are on the order of the stellar radius to obscure a sufficient fraction of
the stellar emission. Specifically, we modeled the flux evolution
as a combination of
(A) the unocculted star, (B) the upper layers of the disk characterized by 
reddening according to the \citet{Cardelli_1989}  extinction law, and 
(C) gray absorption caused by the lower disk layers, i.e.,
\begin{equation}
\text{Observed Spec.} = S_A F_\text{direct} + S_B F_\text{reddened} + S_C F_0 \,, 
\end{equation}
where $S_A$, $S_B$, and $S_C$ are the fractional visible surface areas of the regions A, B, and C from
Fig.~\ref{fig:sketch}, and $F_\text{direct}$, $F_\text{reddened}$, and $F_0$ are the 
direct, reddened, extinguished ($F_0 = 0$) spectra, respectively. We require that $S_A + S_B + S_C = 1$.
This model is intended
as a simplified representation of the warp, which likely 
has a continuous structure. 
We allowed $R_V$ to vary, but find that it's value is consistent with the canonical value of 3.1; and therefore decided to 
fix it to  $R_V=3.1$ in the following. The resulting fit parameters are listed in Table~\ref{tab:fit}. 
The fit shows that the
\begin{itemize}
\item fraction of the stellar surface covered by the 
extra absorber increases   (38 and 64\,\% covering 
      fractions),
\item directly viewed stellar surface decreases (54 and 23\,\%), 
\item reddening, $E(B-V)$, decreases slightly,
\item and that the gray extinction increases (8 and 13\,\% 
      covering fractions)
\end{itemize}
with decreasing flux.
Thus, the flux evolution can be well explained 
by a warp rotating into view and partially occulting 
the stellar photosphere. 

Assuming that the 20th Jan. spectrum represents 
the uneclipsed star,
we can transfer these values to physical lengths  
using the geomtrical relations between inclination $i$, disk warp height $h$, inner 
disk radius $r$, and stellar 
radius $R_\star$ as shown in
Fig.~\ref{fig:sketch}. Specifically, we
used an inclination of about $80^\circ$ and 
$6\,R_\star = 10^{12}\,$cm for the 
radius of the warp \citep[estimated from the dip period and using 
$R_\star \approx 2.37\,R_\odot$, see][]{Fonseca_2014}.
The opaque part of the warp extends 0.37 and 0.26\,$R_\star$ above the 
midplane (for the Jan. 18. and 19. spectra, resp.) while the part of
the warp causing reddening extends $1.07$ and $0.68\,R_\star$ above 
the disk midplane (for the Jan. 18. and 19. spectra, resp.). The CoRoT 
light curves indicate that the warp blocks part of the star 
during about half of the rotation period, which points to a ratio of roughly 3:1
between the azimuthal and vertical dimensions of the warp given the above size estimate. 
We note that, with the viewing geometry derived by \citet{Fonseca_2014}, the 
star is unobscured only during phases when the disk does not significantly
extend above the disk midplane.  

\begin{figure}
\centering
\includegraphics[width=0.49\textwidth]{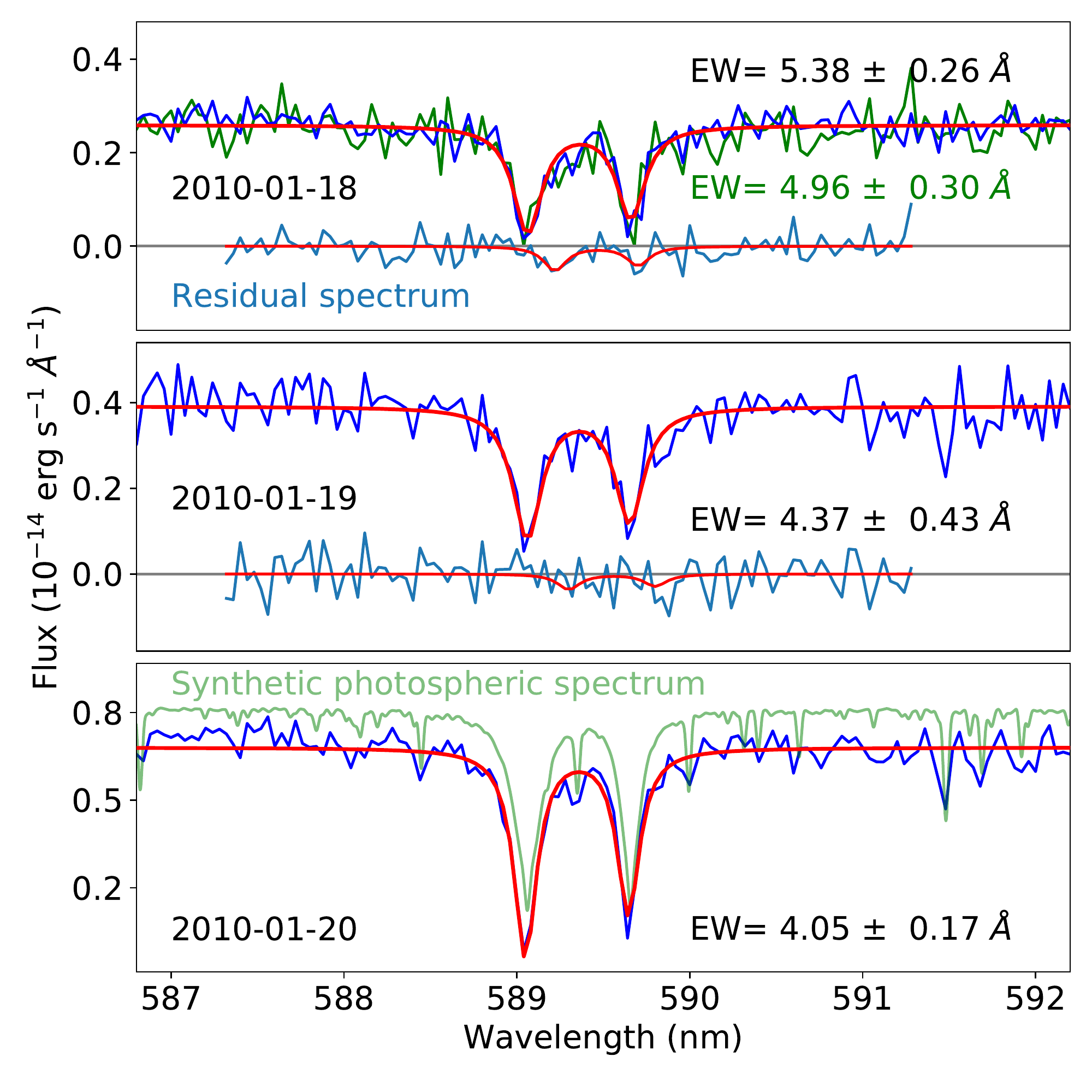}
\includegraphics[width=0.49\textwidth]{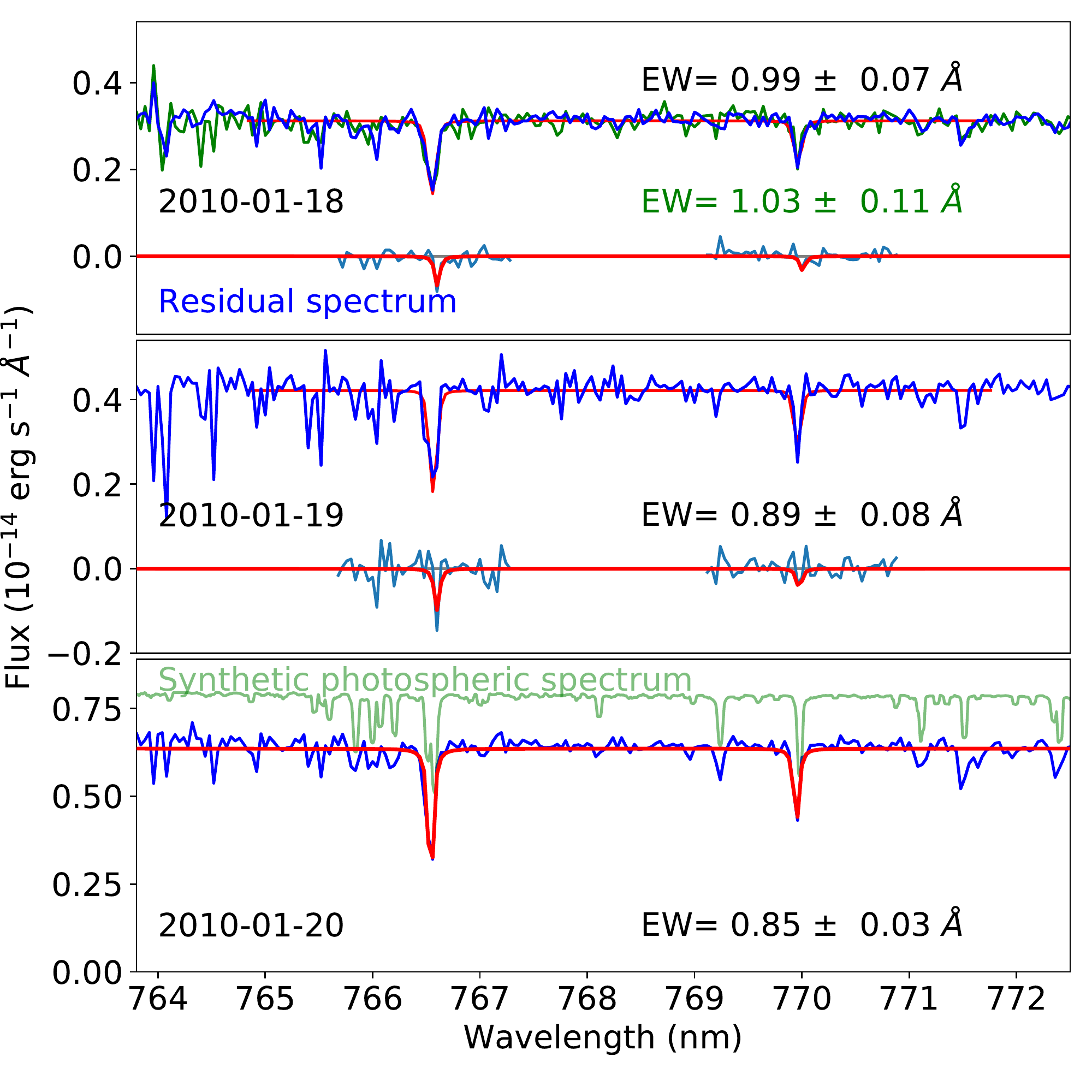}
\caption{{\bf Top}: Na absorption lines. {\bf Bottom}: K absorption lines.\label{fig:Na} 
The photospheric spectrum is based on the PHOENIX model described in the text. The residual 
spectrum is the spectrum minus the brightest spectrum (2010-01-20, the brightest spectrum
was normalized to the same continuum level prior to subtraction, the two 2010-01-18 spectra 
were averaged prior to subtraction). The fit is shown in red (only for the longer 
spectrum from 20th Jan. for clarity).}
\end{figure}

\begin{figure}
\centering
\includegraphics[width=0.49\textwidth]{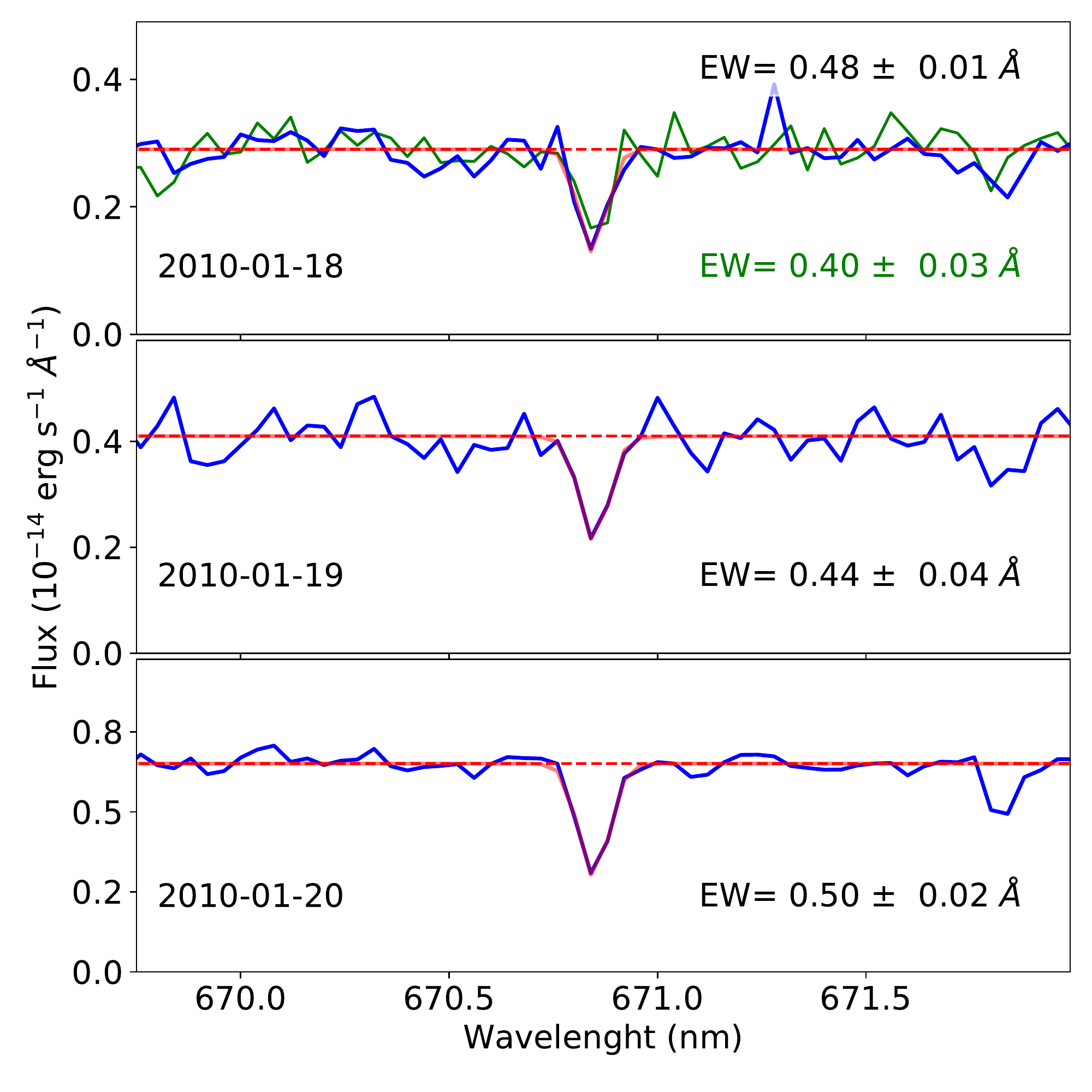}
\caption{Li absorption line.  The fit is shown in red (only for the longer 
spectrum from 20th Jan. for clarity). \label{fig:Li}}
\end{figure}

\begin{table}[t!]
    \centering
    \setlength{\tabcolsep}{0.1cm}
    \caption{Summary of measured line properties (EWs for 2010-01-18 are error-weighted averages).\label{tab:gas}}
    \vspace*{-0.2cm}
    \begin{tabular}{ l l c c c}
    \hline\hline
    Line & & \multicolumn{3}{c}{EW ($\AA$)}\\
         & (nm) & 2010-01-18 & 2010-01-19 & 2010-01-20\\
    \hline
      H$\alpha$ & 656 & $-5.90 \pm 0.18$ & $-6.57 \pm 0.23$ & $-1.88\pm0.46$\\
      Na & 589+589.6 & $5.24\pm0.24$ & $4.37\pm0.43$ & $4.05\pm0.17$\\
      K & 766+770 & $1.00\pm0.08$  &$0.89\pm0.08$ & $0.85\pm0.03$\\
      Li & 671  & $0.47 \pm0.02$ & $0.44\pm0.04$ & $0.50\pm0.02$\\
      He~{\sc i}\tablefootmark{a} & 1083 &  $5.01\pm0.80$ & $4.11\pm0.67$ & $4.39\pm0.56$ \\
    \hline
    \end{tabular}
    \tablefoot{
    \tablefoottext{a}{Absorption component only. The integrated He~{\sc i} EW is provided in Fig.~\ref{fig:Ha}.}  
  }
\end{table}

\subsection{Gas absorption}
The X-Shooter spectra also contain atomic gas absorption 
lines that include nonstellar contributions,
which we later correlate with the dust extinction. 
The spectral resolution of
X-Shooter renders it challenging to differentiate 
interstellar (expected to be narrow) from circumstellar absorption (expected
to be broad) directly in the spectra. Hence, we 
concentrated on the evolution of the observed EW, because 
any ``truly'' interstellar
\citep[e.g.,][]{Pascucci_2015}
or stellar absorption is expected to stay constant.
In this situation, observable changes in the absorption 
require that 
inter- and circumstellar column densities
have roughly similar values.
Circumstellar column density changes  that are small compared
to the interstellar absorption would remain unnoticed.
Conversely, observing EW changes directly implies that 
the circumstellar column densities are sufficiently large.

\subsubsection{Measured line properties}
The Na and K optical doublets are the strongest absorption lines 
in our spectra (Fig.~\ref{fig:Na}). They are common interstellar
absorption lines \citep[e.g.,][]{Welty_2001}, but can also have circumstellar
contributions.
In fact, accretion emission can also contribute to
these lines, but the lack of strong accretion 
signatures\footnote{The H$\alpha$ flux indicates $\log L_{acc}/L_\odot<-2.5$
and $\log L(Na I)/L_\odot=-5.5$.} implies
that this would affect the EW by $\lesssim10\,$m\AA{}.
Similarly, absorption by the accretion funnels has
been suggested to cause broad absorption features in CTTS, mainly
in the strongly accreting ones. Hence, we regard a funnel contribution
to the Na and K absorption insignificant for V\,345~Mon due 
to its very low accretion rate.

In all three
epochs, we clearly detect absorption in the Na and K optical doublets  
while the Na UV lines are not significantly detected
with a combined EW upper limits of 0.7\,\AA{}.
The depth of the Na and K doublets is mainly photospheric, the PHOENIX
model \citep[$T_{eff} = 4900\,$K, $\log\,g=4.50$, solar metallicity, see ][]{Husser_2013}
gives stellar EWs of 3.6\,\AA{} (Na) and 0.6\,\AA{} (K). Neverthless, 
material between us and the photosphere causes extra and variable
absorption since the measured EW is higher than the photospheric value
for both doublets during all epochs.
With the limited spectral resolution of X-Shooter, we cannot 
localize the gas that causes this extra absorption, for example, through its 
velocity structure. However, some of this extra absorption is likely 
interestellar and thus constant between the observing nights 
\citep{Pascucci_2015}. As we regard the accretion process in V\,354~Mon to
have an insignificant effect on the Na and K lines, the variable gas absorption
is likely associated with material in the (inner) disk region.
 
We provide the EWs in Table~\ref{tab:gas} and note that 
the fitted line widths as well as 
the line ratios (measured mean $1.2\pm0.1$ for Na and $2.0\pm0.2$ for K)
are similar 
in all epochs. We are particularly interested in the EW change, because it is related to 
variable amounts of circumstellar material. We denote this value as $\Delta EW$.
Specifically, we find
an increase in 
Na EW of $\Delta EW = 1.2\pm0.3\,$\AA{} (for 2010-01-18 $\rightarrow$ 2010-01-20)
and $\Delta EW = 0.3\pm0.4$\,AA{} (2010-01-19 $\rightarrow$ 2010-01-20). The increases 
in K EWs are $0.15\pm0.08$\,\AA{} (for 2010-01-18
$\rightarrow$ 2010-01-20)  and $0.04\pm0.08$\,\AA{} (for 2010-01-19 $\rightarrow$ 2010-01-20), 
respectively. The data are insufficient to detect a deviation of the blue-to-red line ratio between 
the epochs.

\subsubsection{Column densities}
The conversion from $\Delta EW$ to column density depends 
on the properties of the absorbing material like temperature
and ionization fraction. Specifically for V\,354~Mon, the conversion also
depends on the covering fraction of the absorbing layer since the 
absorber does not cover the entire stellar surface.
Correcting for the fact that the absorber covers only 65\,\% of the visible  
stellar surface (2010-01-18), we
find that circumstellar material causes up to $\Delta EW=1.8\pm0.5\,$\AA{} 
(Na) and $0.2\pm0.1$\,\AA{} (K).

We now consider the properties of the absorbing material
assuming that the same gas causes the Na and K absorption.
At a radius of 
0.1\,au, the gas temperature is expected to be about
1400\,K (equivalent to 0.1\,eV) so that both, Na and K, are neither significantly thermally 
ionized 
(the energies of the first ionization potentials  are 4.3\,eV and 5.1\,eV for Na and K, resp.)
nor is the upper level significantly thermally populated (excitation energies 
are 2.1\,eV and 1.6\,eV, for Na and K respectively). 

Photoionization, however, can cause non-negligible ionization in parts of the innermost
fraction of the disk rim. While a detailed photo-chemical simulation is
beyond the scope of this paper, we ran several simple
cloudy simulations \citep{Ferland_2013} to estimate the impact of photoionization.
Specifically, we reconstructed the stellar emission from available observations across the electromagnetic spectrum as detailed
in Appendix~\ref{sect:input_spectrum}. We used a cylindrical geometry with height and radial extent of 0.1\,au,
a fixed temperature of 1400\,K, and densities such that they reproduce the 
observed extinction ($\log n\sim8$, grain \hbox{abundance $\gtrsim10\times$ISM}). These experiments suggest that photoionization 
might cause a volume averaged ionization somewhere between 
10 and about 90 percent. In the following, we present values assuming no 
photoionization noting that these
are lower limits and the total Na and K column densities might be up to a factor 
of ten higher.

We now continue estimating the Na and K column densities.
Optical depth effects are less pronounced for K.
The line ratio of about two for the K doublet\footnote{Strictly speaking, 
this applies only for the combination of photospheric, inter-, and circumstellar
absorption. However, we do not find indications for a change in the line ratio for 
the difference spectrum.} suggests 
$N_{K}=1.1\times10^{12}\,$cm$^{-2}$  based on 
$\Delta$\,EW $=0.2$\,\AA{}. 
For Na, we assumed that the measured line ratio of 1.3 holds also for the
extra circumstellar absorber and use Table~2 in \citet{Stroemgren_1948}
to find $N_{Na}\approx3\times10^{13}$\,cm$^{-2}$, surprisingly close
to the value expected based on the K column density and their elemental 
abundance ratio.
Both values correspond to 
$N_H \approx 1\dots2\times10^{19}$\,cm$^{-2}$ if Na and K in the disk warp are neutral. 

To provide a rough estimate of the gas density in the disk warp located at $\sim0.1$\,au, we assumed
that the length of the sight-line through the warp is 0.1\,au. Assuming 
a Keplerian disk, differential rotation would quickly
shear apart any structure that extends over substantially different radii unless the structure is maintained or
supported by other processes\footnote{This likely also holds if the warp is caused by the interaction of the disk 
with the stellar magnetic field as the magnetic field strength decreases with $r^3$, i.e.,
the field strength at 0.1\,au is almost a magnitude higher than at 0.2\,au.}. This gives a density of $n_H\sim10^{7}\,$cm$^{-3}$ in the warp.
This value cannot be easily converted to a disk surface density since we do not probe 
the  part of the disk causing the opaque obscuration, where the dust density is higher.

\subsection{Dust vs. gas}
The ratio of gas to dust absorption 
is $\Delta N_H = 1\dots2\times10^{19}\,A_V^{-1}$ (we 
transferred the modeled $E(B-V)$ to $A_V$ using $R_V=3.1$).
This is 
a factor of 100 below the interstellar value for 
a gas-to-dust ratio of 100:1 and implies a strong 
overabundance of dust in the warp or a strong depletion
of neutral, atomic Na and K. Considering a possible (photo-) ionization of 
the gas in the disk warp, the true (total) gas-to-dust ratio might be 
as low as about 10:1, 
which still implies a dust overabundance albeit less severe than 
in the situation assuming dominantly neutral Na and K.

\section{Discussion \label{sect:disc}}
We find large differences between the spectra 
of V\,354~Mon obtained in three consecutive nights. 
The broadband evolution of the SED is dominated by dust extinction. A combination 
of light from the unocculted star, from a reddened stellar spectrum, and 
gray extinction accurately describes the flux evolution for wavelengths 
between 300 and 2000\,nm. Specifically, we find that with decreasing flux  both 
the fraction of the star obscured by the gray absorber as well as the 
fraction of the star subject to reddening increase while the fraction
of the star that remains unobscured decreases.

We also find an evolution in atomic absorption lines with the 
dimmest spectrum exhibiting the largest EW. The variability amounts
to a quarter of the total EW, which we ascribe to circumstellar gas.
The change in the neutral gas column densities is small compared 
to expectations based on interstellar conditions, and photo-ionization
is not likely to make up for the missing gas.
This raises the question
as to whether our analysis strategy misses significant amounts of gas
or overestimates the intervening dust and we consider (1) the doublet's line ratio, (2), elemental abundances, and 
(3) an underestimate of the ionization as possible causes for a low gas 
column density in the following. 

First, we used the doublet's measured line ratios to estimate the increase in 
gas column 
density. However, the extra absorber might have a different, in particular a lower, 
line ratio.
While we do not
find any hint for such a change in the doublet line ratios, the data is 
insufficient to conclusively rule this possibility out.
A ratio closer to unity would
increase the column density corresponding to the EW increase. 
Based on the 
calculations by \citet{Stroemgren_1948}, we expect that this could
increase the column density by about one order of magnitude. 
This would still be compatible with our upper limit from the 
low S/N  Na UV absorption lines, which have
oscillator strengths about a factor of 20 lower than the optical lines. The value of $EW\lesssim0.7\,$\AA{} for the sum of 
both UV lines corresponds to an upper limit in the range of
$\log N_{Na} \lesssim 15$ using the values by \citet[][the ratio between the optical and UV lines reduces to less than ten for 
strongly saturated optical lines]{Stroemgren_1948}.
Higher S/N data with higher spectral resolution would allow one to confirm or rule out changes in the 
line ratio as well as to better constrain the gas column density using 
the UV lines.

Second, the elemental abundances (of the circumstellar material) might
differ from the Sun's photospheric abundances. It is difficult to assess
the error introduced by this effect as abundance patterns for circumstellar 
disks have not been measured. Studies of photospheric abundances in 
Herbig stars suggest a depletion in refractory elements,
which has been related to dust filtering by planet formation within the 
protoplanetary disk \citep{Kama_2015}.
Extending this argument to volatile elements such as Na and K suggests that
they could be overabundant so that a gas-to-dust ratio based on Na and K absorption
would appear erroneously low. 
In principle, other species can be measured using, for example, UV transmission spectroscopy (in particular H), however,
V\,354~Mon is too dim  to apply this method. Therefore, 
it is desirable to identify nearby
systems with a stable AA~Tau-like phenomenon and a (very) low accretion rate
to avoid effects by accretion variability. 

Third, our simple geometry used for  calculating the photo-ionization
might underestimate the ionization state of the plasma and 
a detailed model is needed to assess this possibility. We expect, however,
that a more realistic structure rather reduces the ionization, because the
high-energy photons would be efficiently absorbed in an outer, low density layer.
Hence, fewer high-energy photons would reach the high density region in which 
the ionization conditions are closer to thermodynamic equilibrium. In total, we expect this to result in a lower average ionization.

Another possibility is that we overestimate the dust mass. However, our
method is most prone to miss any large dust grains that do not
cause reddening nor a significant dimming, and unlikely misses smaller dust
that causes the optical reddening. 
Therefore, our dust estimate
can be considered a lower limit so that our analysis strongly suggests
a low gas-to-dust ratio in the disk warp, about 1:1. Factoring in 
the uncertainty in the gas column density, we still find strongly 
gas depleted material (perhaps up to roughly 10:1 gas-to-dust compared to
100:1 in the ISM). Such a gas depletion appears not unreasonable 
given the low accretion rate of V\,354~Mon. Perhaps the disk is already
evolving toward a debris disk and the lack of a sufficient reservoir of
gas in the inner disk also causes the accretion rate to decrease.

\section{Conclusions \label{sect:concl}}

Our analysis demonstrates that flux calibrated, broadband data
such as X-Shooter spectra allow us to simultaneously determine
the dust and gas properties. These properties, when combined with sufficient
time sampling, provide access to the gas-to-dust ratio of 
disk warps that periodically partially occult the central star.
We also describe how higher S/N data combined with  better phase 
sampling can overcome limitations in the data analyzed here, which were
obtained for a different purpose. Also, better phase sampling could reveal
any phase differences between the dust and the gas as well as 
potentially allow a refined modeling of the height structure of
the disk warp compared to our toy model with only three zones.

We find a gas-to-dust ratio for the inner disk regions around V\,354~Mon 
that is a factor of ten or more lower than typical ISM values. 
One possibility is
that V\,354~Mon's disk is already evolving toward a debris disk, 
which could also be related to the low stellar accretion rate. 
Another mechanism possibly resulting in a  low gas-to-dust ratio in the inner 
disk is the radial drift of large particles from the outer disk, which then fragment in the inner regions due to the high relative velocities \citep[e.g.,][]{Birnstiel_2010}.
Low gas-to-dust ratios in the inner disk might indicate that the streaming instability can be triggered in the proximity of the central star, leading to the formation of rocky cores at \hbox{the $\lesssim$\,au} scale.

Better characterizing the properties of inner disk warps is also 
relevant in view of recent discoveries of 
shadows seen in very high spatial resolution scattered light images
\citep[e.g.,][for disks around HAeBe stars]{Marino_2015,Stolker_2016,Juhasz_2017} 
or as rotating structures observed with HST, for example, for the nearby CTTSs TW~Hya \citep{Debes_2017}.
An improved height model of the inner disk warp could
address whether shadows seen in images of stellar light reflected 
at the disk surface might be related to the same disk warps that we
studied in this paper.
Combining complementary constraints for inner disk properties will eventually lead to 
a better understanding of the physical conditions in the terrestrial planet forming region.

\begin{acknowledgement}
PCS thanks Jerome Bouvier for interesting discussions and very valuable comments.
Based on observations made with ESO Telescopes at the La Silla 
Paranal Observatory under programme ID 084.C-1095(A). PCS acknowledges support
by DLR 50 OR 1307 and DFG SFB 676. PCS and CFM gratefully acknowledge ESA Research Fellowships, during which significant parts of the work were performed.
DF acknowledges support from the Italian Ministry of Education, Universities and Research project SIR (RBSI14ZRHR).  Support for HMG was provided by the National Aeronautics and Space 
Administration through Chandra Award Number GO6-17021X issued by the Chandra X-ray 
Observatory Center, which is operated by the Smithsonian Astrophysical 
Observatory for and on behalf of the National Aeronautics Space Administration 
under contract NAS8-03060.
\end{acknowledgement}

\bibliographystyle{aa}
\bibliography{v354}

\begin{appendix}
\section{Veiling}
\begin{figure}
\includegraphics[width=0.45\textwidth]{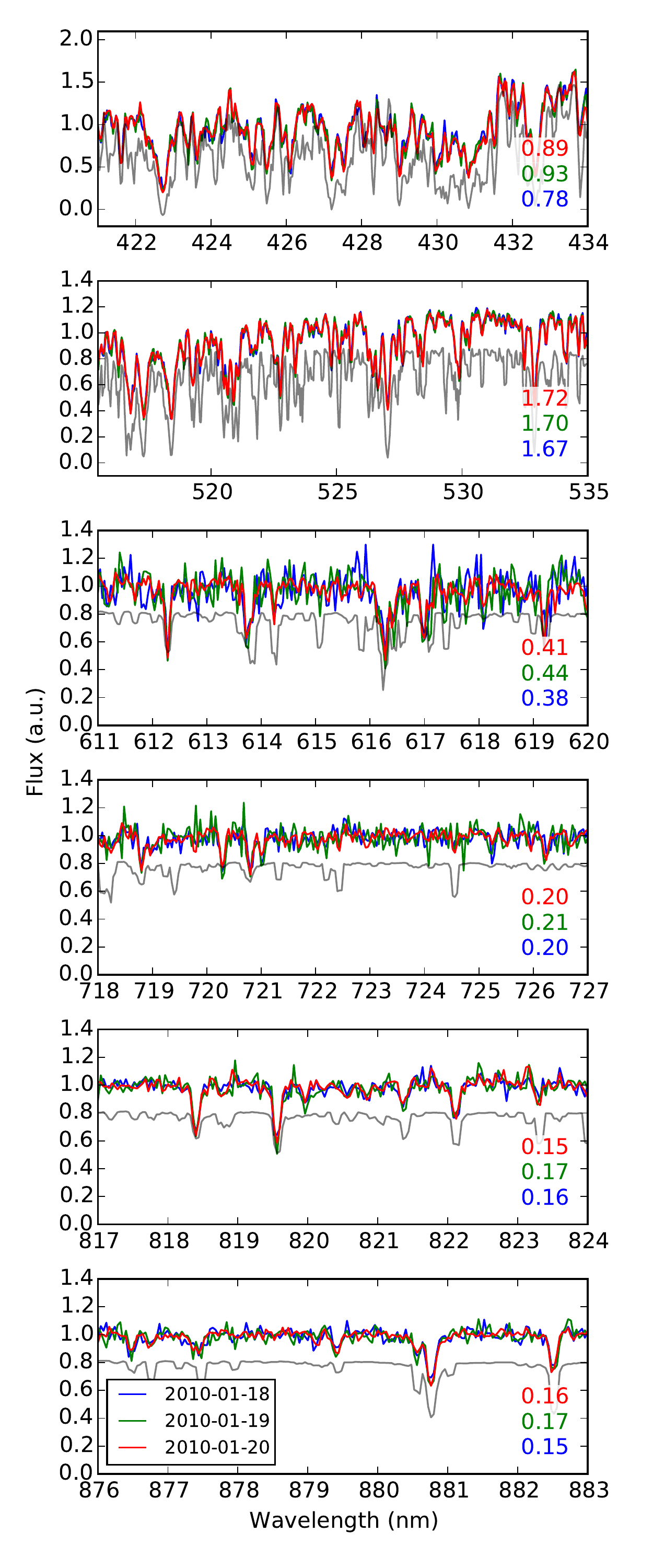}
\caption{Stellar features at different wavelengths accross the spectral
range of X-Shooter. A synthetic spectrum is overlayed in gray but offset by 0.2 for 
clarity. Numbers indicate the pseudo equivalent width calculated 
over the displayed wavelength region.   \label{fig:veiling}}
\end{figure}

We inspected the line depth in several wavelength regions accros the optical 
spectrum. We then calculated a pseudo equivalent width as 
\begin{equation}
  \int_{\lambda_0}^{\lambda_1} 1 - F_{norm}(\lambda) d\lambda\,.
\end{equation}
These pseudo equivalents are provided in Fig.~\ref{fig:veiling}.

\begin{figure}
\centering
\includegraphics[width=0.49\textwidth]{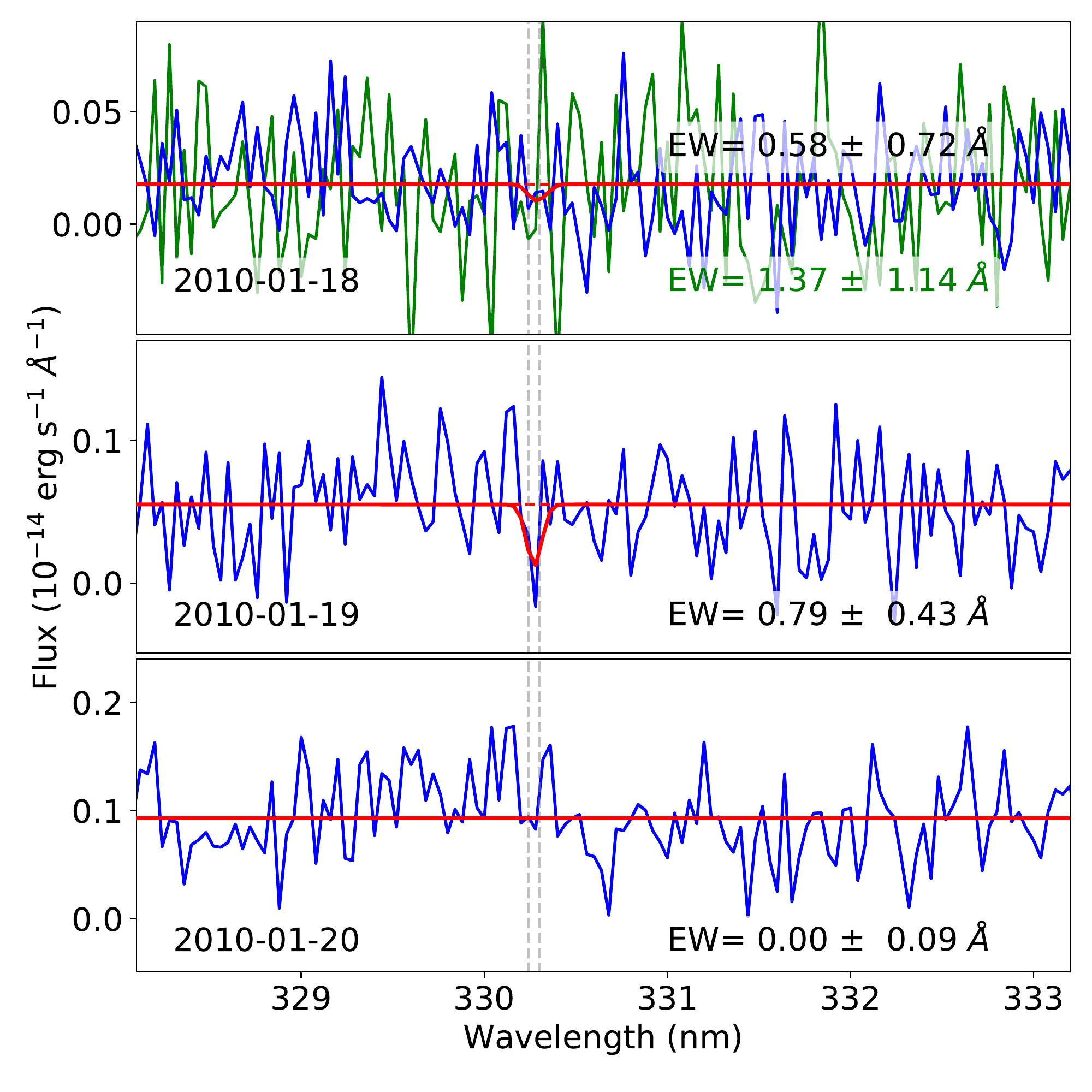}
\caption{Na UV absorption lines with the locations of the lines plotted. The red line indicates the best fit model (to the longer spectrum for 2010-01-18).\label{fig:Na_UV}}
\end{figure}

\section{Input spectrum for \texttt{cloudy} simulations \label{sect:input_spectrum}}
The cloudy simulations need a reasonable input spectrum that includes
the high-energy portion of the stellar spectrum. We constructed such 
a spectrum based on, first, the flux calibrated X-Shooter spectrum for 
the optical part, second, the RECX~11 FUV spectrum from \citet[][RECX~11 is a weakly accreting K4 similar to V~354~Mon]{France_2014},
and third, X-ray data published by \citet[][$\log L_X=30.61$]{Dahm_2007}.
We then interpolated a solar spectrum to fill the EUV part of the spectrum
using a first order polynomial fixed at the two connecting points (X-ray/FUV).
The long wavelength part ($\lambda\gg1\,\mu$m) of the spectrum was approximated 
by a blackbody. We experimented with several normalizations but found the effects
to be on the same order as mild changes in the (unknown) gas-to-dust ratio of 
the inner disk rim. We therefore used $L=4\times10^{33}\,$erg\,s$^{-1}$, which matches
the flux calibrated X-Shooter spectrum.

\end{appendix}
\end{document}